\documentclass[a4paper,USenglish,cleveref, autoref, thm-restate]{lipics-v2019}

\usepackage{tikz}
\usetikzlibrary{patterns}
\title{Reducing Moser's Square Packing Problem to a Bounded Number of Squares} 

\author{Meike Neuwohner}{Research Institute for Discrete Mathematics, Universit\"at Bonn}{neuwohner@or.uni-bonn.de}{https://orcid.org/0000-0002-3664-3687}{}

\authorrunning{M. Neuwohner} 

\Copyright{Meike Neuwohner} 

\ccsdesc[500]{Theory of computation~Computational geometry}

\keywords{Square packing, Moser, reduction to a bounded number of squares} 

\category{} 

\relatedversion{} 

\supplement{}

\acknowledgements{I would like to thank Prof. Dr. Stefan Hougardy for the idea to study the whitespace arising in square packings.}

\nolinenumbers 

\hideLIPIcs  

\begin{document}

\maketitle

\begin{abstract}
The problem widely known as \emph{Moser's Square Packing Problem} asks for the smallest area $A$ such that for any set $S$ of squares of total area $1$, there exists a rectangle $R$ of area $A$ into which the squares in $S$ permit an internally-disjoint, axis-parallel packing. It was formulated by Moser~\cite{MoserProblemFormulation} in 1966 and remains unsolved so far. The best known lower bound of $\frac{2+\sqrt{3}}{3}\leq A$ is due to  Novotn\'y~\cite{Novotny95} and has been shown to be sufficient for up to $11$ squares by Platz~\cite{Platz}, while Hougardy~\cite{Hougardy11} and Ilhan~\cite{Ilhan} have established that $A<1.37$.\\ In this paper, we reduce Moser's Square Packing Problem to a problem on a finite set of squares in the following sense: We show how to compute a natural number $N$ such that it is enough to determine the value of $A$ for sets containing at most $N$ squares with total area 1.
\end{abstract}
\newpage
\section{Introduction}
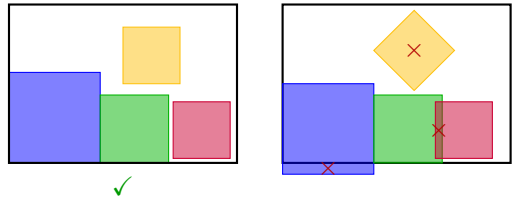
\begin{figure}[t]
 \centering
 	\begin{tikzpicture}[scale = 0.3]
 	\draw[draw = blue, fill = blue, fill opacity = 0.5] (0,0) rectangle (4,4);
 	\draw[draw = green!70!black, fill = green!70!black, fill opacity = 0.5] (4,0) rectangle (7,3);
 	\draw[draw = yellow!50!orange, fill = yellow!50!orange, fill opacity = 0.5] (5,3.5) rectangle (7.5,6);
 	\draw[draw = red!80!blue, fill = red!80!blue, fill opacity = 0.5] (7.2,0.2) rectangle (9.7,2.7);
 	\draw[draw = black, thick, fill = none] (0,0) rectangle (10,7);
 	\node at (5,-1){\textcolor{green!60!black}{$\checkmark$}};
 	\begin{scope}[shift = {(12,0)}]
 	\draw[draw = black, thick, fill = none] (0,0) rectangle (10,7);
 	\draw[draw = blue, fill = blue, fill opacity = 0.5] (0,-0.5) rectangle (4,3.5);
 	\draw[draw = green!70!black, fill = green!70!black, fill opacity = 0.5] (4,0) rectangle (7,3);
 	\draw[draw = red!80!blue, fill = red!80!blue, fill opacity = 0.5] (6.7,0.2) rectangle (9.2,2.7);
 	\draw[draw = yellow!50!orange, fill = yellow!50!orange, fill opacity = 0.5] (5.77,3.2)--(7.54,4.97)--(5.77,6.74)--(4,4.97)--(5.77,3.2);
 	\node at (5.77,4.97) {\textcolor{red!70!black}{$\times$}};
 	\node at (6.85,1.425) {\textcolor{red!70!black}{$\times$}};
 	\node at (2,-0.25) {\textcolor{red!70!black}{$\times$}};
 	\end{scope}
 	\end{tikzpicture}
 	\caption{A feasible (left) and an infeasible (right) packing.}\label{FigPackingMoser}
 \end{figure}

\begin{figure}[t]
  \centering
 		\begin{tikzpicture}[scale = 3]
 			\draw[draw = blue, thick, fill = blue, fill opacity = 0.5] (0,0) rectangle (0.7071,0.7071);
 			\draw[draw = green!70!black, thick, fill = green!70!black, fill opacity = 0.5] (0.7071, 0) rectangle (1.1153,0.4082);
 			\draw[draw = green!70!black, thick, fill = green!70!black, fill opacity = 0.5] (0.7071,0.4082) rectangle (1.1153,0.8164 );
 			\draw[draw = green!70!black, thick, fill = green!70!black, fill opacity = 0.5] (1.1153, 0) rectangle (1.5235,0.4082);
 			\draw[draw = black, thick, fill = none] (0,0) rectangle (1.5235,0.8164);
 			\draw[<-, black, thick] (-0.15, 0)--(-0.15, 0.26);
 			\draw[->, black, thick] (-0.15, 0.5564)--(-0.15, 0.8164);
 			\node at(-0.15,0.4082) {\small$\frac{2}{\sqrt{6}}$};
 			\draw[<-, black, thick] (0, -0.15)--(0.46175, -0.15);
 			\draw[->, black, thick] (1.06175, -0.15)--(1.5235, -0.15);
 			\node at(0.76175,-0.15) {\small$\frac{1}{\sqrt{2}}+\frac{2}{\sqrt{6}}$};
 			\begin{scope}[shift = {(3,0)}]
 			\node at (0,0.27){\textcolor{green!70!black}{$s_2=s_3=s_4=\frac{1}{\sqrt{6}}$}};
 			\node at (0,0.54){\textcolor{blue}{$s_1=\frac{1}{\sqrt{2}}$}};
 			\end{scope}
 			\end{tikzpicture}
 			\caption{The lower bound example provided by Novotný.}\label{FigWorstCase}
 	\end{figure}
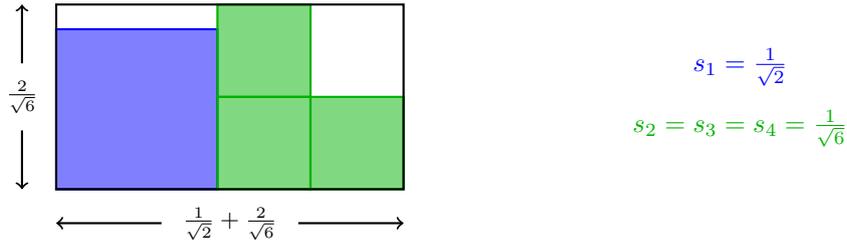

 In 1966, Leo Moser \cite{MoserProblemFormulation} asked for the minimum area $A$ such that for any (possibly infinite) set $S$ of squares of total area $1$, there exists a rectangle $R$ of area $A$ and an axis-parallel, internally disjoint packing of $S$ into $R$ (see Figure~\ref{FigPackingMoser}). The fact that the problem under consideration indeed produces a minimum instead of just an infimum follows from a result by Martin~\cite{martin2002compactness}.\\
 The first bounds on the value of the area $A$ asked for in the problem definition were provided by Moon and Moser \cite{MoonMoser} in 1967, showing that $1.2<A\leq 2$. While the lower bound arises by studying the case of two squares, the upper bound can be proven by considering a packing that accommodates the squares in horizontal layers in order of magnitude. The latter idea was generalized by Meir and Moser \cite{MeirMoser} to prove that for any dimension $k>0$, there exists an axis-parallel, internally disjoint packing of a (possibly infinite) set of $k$-dimensional hypercubes with edge lengths given by $x_1\geq x_2\geq \dots$ and finite total volume $V$ into a rectangular parallelepiped with edge lengths $a_1,\dots,a_k$ satisfying $a_j\geq x_1$ for all $j=1\dots,k$ as well as \begin{equation}
 x_1^k+\Pi_{j=1}^k (a_j-x_1)\geq V. \label{EqMeirMoser}
 \end{equation}The special case $k=2$ has found several applications in subsequent works, see for example \cite{Hougardy11} or \cite{Novotny96}.\\
 In 1970, Kleitman and Krieger~\cite{KleitmanKrieger70} proved that any set of squares of total area $1$ can be packed into a rectangle with edge lengths $1$ and $\sqrt{3}$, which in particular implies an upper bound of $A\leq \sqrt{3}<1.733$. Moreover, five years later, they showed that a rectangle of edge lengths $\frac{2}{\sqrt{3}}$ and $\sqrt{2}$  (and area $\frac{4}{\sqrt{6}}<1.633$) can accommodate any \emph{finite} set of squares of total area $1$ \cite{KleitmanKrieger75}, which was generalized to the case of possibly infinite families of squares by Zernisch~\cite{Zernisch12} in 2012.\\
 The best known lower bound of $\frac{2+\sqrt{3}}{3}$, which is actually believed to be the correct answer to the problem, was provided by Novotný~\cite{Novotny95} in 1995, who considered an instance consisting of one square of area $s_1^2=\frac{1}{2}$ and three squares of area $s_2^2=s_3^2=s_4^2=\frac{1}{6}$ (see Figure~\ref{FigWorstCase}). For packing up to $5$ squares, he proved that an area of $\frac{2+\sqrt{3}}{3}$ indeed suffices \cite{Novotny99}, which was extended to up to $11$ squares by Platz~\cite{Platz}. Moreover, using the results obtained in \cite{MeirMoser} and \cite{KleitmanKrieger75}, Novotný~\cite{Novotny96} proved an upper bound of $A< 1.53$ for the general case of possibly infinitely many squares.\\
 The best known upper bound of $A<1.37$ is due to Hougardy~\cite{Hougardy11} and Ilhan~\cite{Ilhan}. They employed a computer-aided case distinction, where various packing routines are applied to accommodate the largest first squares and a rectangle meeting the criterion \eqref{EqMeirMoser} containing the remaining ones. In doing so, Hougardy~\cite{Hougardy11} first proved an upper bound of $A<1.40$, which was then refined to first $1.39$ and later $1.37$ in cooperation with Ilhan~\cite{Ilhan} by using a different set of packing methods.\\
 In this paper, we present a general approach that allows us to reduce the task of showing that any set of squares of total area $1$ can be packed into a rectangle of area $\frac{2+\sqrt{3}}{3}\leq F\leq 1.37$ to the problem of proving this statement for \emph{finite} sets containing at most $N$ squares, where $N$ is a natural number that can be easily computed.\\ In a similar spirit, Liu~\cite{Liu17} has shown (among other results) that for each infinite set of squares $S=\{S_1,S_2,\dots\}$ of total area $1$, the minimum area of a rectangle into which the finite subset $\{S_1,S_2,\dots,S_n\}$ of $S$ can be packed axis-parallel and internally disjoint converges to the minimum area of a rectangle able to accommodate all of the squares in $S$ as $n$ approaches $\infty$. This is a relatively straightforward consequence of the Meir-Moser Theorem.\\ However, the strategy applied by Liu can only be used to reduce the infinite problem to a finite one within a certain error rate. More precisely, both \begin{itemize}
 		\item considering a packing of $S_1,\dots,S_n$ into some rectangle $R$ of area $F$ and then slightly inflating $R$ to ensure that an additional rectangle being able to accommodate the remaining squares due to \eqref{EqMeirMoser} can be packed as well or
 	\item packing all of the squares $S_{n},S_{n+1},\dots$ into some larger square $S'_{n}$ meeting the requirements of \eqref{EqMeirMoser}
 	\end{itemize}will inevitably increase the total area of objects to be packed from $1$ to $1+\epsilon$ for some $\epsilon >0$. Therefore, these approaches only allow to deduce an area bound of $F\cdot(1+\epsilon)$ for a packing of a possibly infinite set of squares from an area bound of $F$ established for some finite case.\\
 The same problem also seems to be inherent in the previous attempts to achieve improved upper bounds since most of them rely on the application of \eqref{EqMeirMoser} to get down to a finite number of rectangular objects to be packed.\\
 In order to overcome this difficulty, instead of trying to pack a finite sub-collection of "large" squares together with an additional rectangle containing the "small" ones, we consider a packing that only respects the "large" squares and then show how to fit the remaining squares into the arising \emph{whitespace}. The latter is defined to be the closure of the set of points in the enclosing rectangle that are not covered by any of the squares already packed (see Definition~\ref{DefWhitespace}).\\ Note that a similar argument also appears in \cite{Platz}, proving that an area of $\frac{2+\sqrt{3}}{3}$ suffices to pack sets consisting of at most $11$ squares of total area $1$. In this context, the area of a largest square present in the whitespace arising from a packing of $n$ squares into a rectangle of the desired area $F$ is bounded from below by some constant $a_n$ depending only on $n$ (and $F$). This allows to inductively reduce the case where the smallest edge length occurring among those of the finitely many squares considered is less than $a_n$ to the case where the number of squares is one less. However, the bounds presented in \cite{Platz} are too weak to be applicable in the sort of argument we aim at.\\
 Therefore, in the following section, we will first see how to produce sufficient bounds on the size of squares that can be accommodated within the whitespace, and then see how to apply these in order to obtain the desired result. 

 \section{Reduction to a Finite Number of Squares\label{SecReductionFinite}}
Let a target area $F\geq \frac{2+\sqrt{3}}{3}$ be given. This section shows how to compute a natural number $N(F)$ with the following property: If for any set of at most $N(F)$ squares of total area $1$, there exists a rectangle of area $F$ into which these squares can be packed axis-parallel and internally disjoint, then the same statement holds for any, possibly infinite set of squares of total area $1$.\\ From the way we compute $N(F)$, it is not hard to see that $N:=\max\{N(F),F\in[\frac{2+\sqrt{3}}{3},1.37]\}$ exists and is attained by $N(\frac{2+\sqrt{3}}{3})>4$. Note that these properties already imply that if we denote the correct answer to Moser's Square Packing Problem (when restricted to finite sets containing at most $N$ squares) by $A$ ($A'$), then $A=A'$:\\ First, $A < 1.37$ by \cite{Ilhan}. Furthermore, we clearly have $A'\leq A$: If for any set of squares of total area $1$, there is a rectangle of area $A$ into which the given squares permit an axis-parallel, internally disjoint packing, then this in particular also holds for any set of \emph{at most $N$} squares of total area $1$. Hence, $A'\leq A<1.37$.\\ Conversely, the fact that $N>4$ additionally implies that $\frac{2+\sqrt{3}}{3}\leq A'$ since the example yielding this lower bound consists of four squares \cite{Novotny95}. But now, by our choice of $N$, we must have $N(A')\leq N$ since $\frac{2+\sqrt{3}}{3}\leq A'<1.37$. As a consequence, by definition of $A'$, we know that for any set of at most $N(A')\leq N$ squares of total area $1$, there is a rectangle of area $A'$ in which the squares can be accommodated. By our choice of $N(A')$, this yields $A\leq A'$. Hence, we obtain $A=A'$ and $N$ possesses the desired property. \\
In order to determine numbers $N(F),F\in[\frac{2+\sqrt{3}}{3},1.37]$ as promised, we first note that without loss of generality, we can assume a set of squares of total area $1$ to be given by a non-increasing sequence $s_1\geq s_2\geq\dots$ of non-negative edge lengths satisfying $\sum_{i=1}^\infty s_i^2 =1$. This is because for any set $S$ of squares of total area $1$, the number of squares in $S$ with edge lengths in $[\frac{1}{n+1},\frac{1}{n}]$ for some fixed positive integer $n$ must be finite, and squares of edge length zero do not influence the packing task. For a given sequence $s_1\geq s_2\geq\dots$, we denote the corresponding squares by $S_1,S_2,\dots$.\\ We further observe that if $s_1$ is small enough, the criterion established by Meir and Moser provides a packing of $S$ into a rectangle of area $F$, so we can assume that $s_1$ is greater than a certain constant. Next, we show that there exists constants $N_0(F)$ and $c(F)$ with the following properties: If the total area of the squares with indices $N_0(F)+1$, $N_0(F)+2$, $\dots$ amounts to more than $c(F)^2$, (implying that the total area of the first $N_0(F)$ largest squares is less than $1-c^2(F)$,) then their edge lengths must be "small" and they can, therefore, be packed "efficiently". On the other hand, if the total area of $S_{N_0(F)+1}$, $S_{N_0(F)+2}$, $\dots$ is at most $c(F)^2$, then we also know that there must be some $N_0(F)+1\leq n\leq \lfloor\mathrm{e}^2\cdot N_0(F)\rfloor$ such that $s_n\leq\frac{c}{\sqrt{n}}$. For such $n$, we show that for any packing of the first $n$ squares into a rectangle of area $F$, the remaining squares can be accommodated in the arising \emph{whitespace} (see Definition~\ref{DefWhitespace}). This implies that it suffices to consider sets of at most \mbox{$N(F):=\lfloor\mathrm{e}^2\cdot N_0(F)\rfloor$} squares to find the correct answer to Moser's Square Packing Problem.\\ The remainder of this section provides the omitted details. 
We first restate the following, already mentioned results by Meir, Moon and Moser because we need them for the subsequent arguments.
 	\begin{theorem}[Moon, Moser \cite{MoonMoser}]
 	Any set of squares of total area $V$ and with maximum occurring edge length $x$ can be packed into any rectangle with edge lengths $a_1$ and $a_2$ satisfying $\min\{a_1,a_2\}\geq x$ and $2\cdot V\leq a_1\cdot a_2$.
 	\label{double_area}
 \end{theorem}
 \begin{theorem}[Meir, Moser \cite{MeirMoser}]
 	Any set of squares with total area $V$ and maximum occurring edge length $x$ can be packed into any rectangle with edge lengths $a_1$ and $a_2$ satisfying $\min\{a_1,a_2\}\geq x$ and $V\leq x^2+(a_1-x)\cdot (a_2-x)$.
 	\label{MeirMoser}
 \end{theorem}
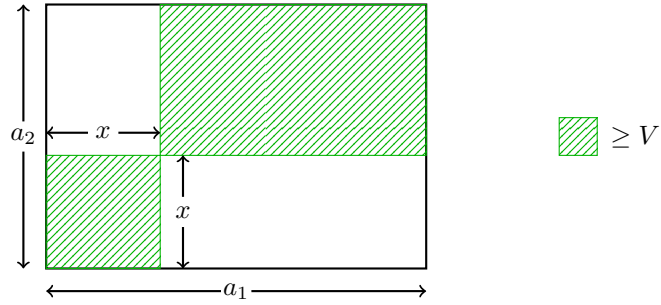
\begin{figure}[t]
\centering
	\begin{tikzpicture}[scale = 0.5]
	\draw[draw = black, thick, fill = none] (0,0) rectangle (10,7);
	\draw[draw=green!70!black, pattern = north east lines, pattern color = green!70!black] (0,0) rectangle (3,3);
	\draw[draw=green!70!black, pattern = north east lines, pattern color = green!70!black] (3,3) rectangle (10,7);
	\draw[thick, black, <-](0,-0.6)--(4.6,-0.6);
	\node at (5,-0.6){$a_1$};
	\draw[thick, black, ->] (5.4,-0.6)--(10,-.6);
	\draw[thick, black, <-] (-0.6,0)--(-0.6,3.1);
	\node at (-0.6,3.5) {$a_2$};
	\draw[thick, black, ->] (-0.6, 3.9)--(-0.6,7);
	\draw[thick, black, <-] (3.6,0)--(3.6,1.1);
	\node at (3.6,1.5) {$x$};
	\draw[thick, black,->] (3.6, 1.9)--(3.6,3);
	\draw[thick, black, <-] (0,3.6)--(1.1, 3.6);
	\node at (1.5, 3.6) {$x$};
	\draw[thick, black, ->] (1.9, 3.6)--(3,3.6);
	\draw[draw=green!70!black, pattern = north east lines, pattern color = green!70!black] (13.5,3) rectangle (14.5,4);
	\node at (15.5,3.5){$\geq V$};
	\end{tikzpicture}
	\caption{Visualization of the condition established by Meir and Moser. The picture does \emph{not} indicate the packing strategy applied to prove the respective result.}
\end{figure}
 \begin{corollary}
 	Let $F > 1$, $V>0$ and $C >0$. Then any set of squares of total area $V$ and with maximum edge length at most $\frac{(F-1)V}{C}$ fits into any rectangle with edge lengths $a_1$ and $a_2$ satisfying $F\cdot V=a_1\cdot a_2$ and $a_1+a_2\leq C$. \label{circumference}
 \end{corollary}
 \begin{proof}
 	Let $x$ denote the maximum edge length among the given squares and let $a_1$ and $a_2$ be as in the statement of the corollary. First, note that \[x\leq \frac{(F-1)V}{C}\leq \frac{FV}{C}\leq \frac{a_1\cdot a_2}{a_1+a_2}\leq \min\{a_1,a_2\}.\] Moreover, $x\cdot(a_1+a_2)\leq \frac{(F-1)V}{C}\cdot C=(F-1)V$. Hence, \[x^2+(a_1-x)\cdot(a_2-x)= 2x^2+a_1\cdot a_2 -x\cdot (a_1+a_2)\geq 2\cdot 0+ FV-(F-1)V=V,\] so Theorem~\ref{MeirMoser} yields the claim.	
 \end{proof}
 Let $F\geq\frac{2+\sqrt{3}}{3}$ be fixed.
 \begin{proposition}
 	Let $(s_i)_{i\in\mathbb{N}^+}$ be a non-increasing sequence of non-negative numbers such that $\sum_{i=1}^{\infty}s_i^2=1$.\newline If $s_1\leq \frac{1}{10}$, then there exists an internally disjoint packing of the squares with edge lengths $(s_i)_{i\in\mathbb{N}^+}$ into a square of edge length $\sqrt{F}$.
 	\label{small_s1}
 \end{proposition}
 \begin{proof}
 	We have $\sqrt{F}\geq\sqrt{\frac{2+\sqrt{3}}{3}}>\frac{11}{10}>\frac{1}{10}\geq s_1$ and, hence, also $\sqrt{F}-s_1 > 1$. Therefore, we get $s_1^2+(\sqrt{F}-s_1)^2>1$, so Theorem~\ref{MeirMoser} yields the claim.
 \end{proof}
 \begin{definition}[whitespace]
  Let a closed rectangle $R$ and a set $\mathcal{S}$ of closed squares be given, such that each of the squares in $\mathcal{S}$ is contained within $R$ and the interiors of two distinct squares from $\mathcal{S}$ do not intersect. We define the whitespace $\mathrm{whitespace}(R,\mathcal{S})$ arising from this packing of the squares in $\mathcal{S}$ into $R$ to equal \[\mathrm{whitespace}(R,\mathcal{S}):=\overline{R\backslash \bigcup_{S\in\mathcal{S}}S},\] where for a set $A$ of points in the Euclidean plane, $\bar{A}$ denotes its topological closure.\label{DefWhitespace}
 \end{definition}
 \begin{lemma}
 	Let $c:= \sqrt{\left(\frac{3}{10}\right)^2+\frac{F-1}{5}}-\frac{3}{10}$. Then for $n\geq\max\{ (10F+\frac{1}{10})^2, 100c^2\}$ and any packing of $n$ squares of total area at most $1$ into a rectangle with edge lengths $\frac{1}{10}\leq W\leq H$ and area $W\cdot H=F$, any set of squares of total area at most $c^2$ and maximum edge length at most $\frac{c}{\sqrt{n}}$ permits an axis-parallel, internally disjoint packing into the arising whitespace. \label{PackRemainingSquares}
 \end{lemma}
 \begin{proof} Let $n\geq \max\{ (10F+\frac{1}{10})^2, 100c^2\}$ be a natural number and let $n$ squares $S_1,\dots,S_n$ with edge lengths $s_1,\dots,s_n$ and total area at most $1$ be given. Consider a packing of these squares into an enclosing rectangle $R$ with edge lengths $\frac{1}{10}\leq W\leq H$ and area $F$.\\ Moreover, let a set of squares of total area at most $c^2$ and maximum edge length at most $\frac{c}{\sqrt{n}}$ be given by a sequence $\frac{c}{\sqrt{n}}\geq s_{n+1}\geq s_{n+2}\geq\dots$. For a finite set, we have $s_j=0$ from some point on.\\
 	Let $k\geq n+1$ and assume that squares of edge lengths $s_{n+1},\dots,s_{k-1}$ have already been accommodated within the whitespace.
 	Our goal is to find a lower bound on the area of the set of points in $R$ that serve as potential midpoints for squares of edge length $s_k$ contained in the remaining whitespace. If we can provide a positive lower bound on the area of the set of potential square midpoints for each $k\geq n+1$ for which $s_k>0$, this in particular implies that this set will never be empty. Hence, we can inductively construct a packing of the squares with edge length $s_{n+1}\geq s_{n+2}\geq\dots$ into the whitespace by always placing the next square at the lexicographically minimum feasible position. The latter exists as the space of feasible positions is compact.\\ Let $k\geq n+1$ such that $s_k>0$ and assume that squares $(S_i)_{i=n+1}^{k-1}$ with edge lengths $(s_i)_{i=n+1}^{k-1}$ have been placed axis-parallel and internally disjoint within the whitespace. Observe that the square of edge length $s_k$ with midpoint $p\in R$ is contained within the (remaining) whitespace if and only if the $l_\infty$-distance of $p$ to each of the $S_i$ for $i=1,\dots,k-1$ and to the boundary of $R$ is at least $\frac{s_k}{2}$. For $i=1,\dots,k-1$, let $S'_i$ denote the square arising from $S_i$ by extending it by $\frac{s_k}{2}$ in positive and negative $x$- and $y$-direction and let $F_i:=S'_i\backslash S_i$ for $i=1,\dots,n$. Moreover, let $R'$ be the rectangle with edge lengths $W-s_k$ an $H-s_k$ lying centered within $R$ (note that $s_k\leq\frac{c}{\sqrt{n}}\leq\frac{1}{10}\leq W\leq H$ by our choice of $n$) and let further $F_R:=R\backslash R'$. Then, for a point $p$ openly contained in the original whitespace, the square with midpoint $p$ and edge length $s_k$ is contained in the remaining whitespace after additionally packing $(S_i)_{i=n+1}^{k-1}$ if and only if $p$ is not openly contained within any of the frames $(F_i)_{i=1}^n$ around the squares $(S_i)_{i=1}^{n}$, any of the squares $(S'_i)_{i=n+1}^{k-1}$, or the frame $F_R$ inscribed into $R$. Figure~\ref{FigFeasibleMidpoints} illustrates the case $k=n+1$.
 	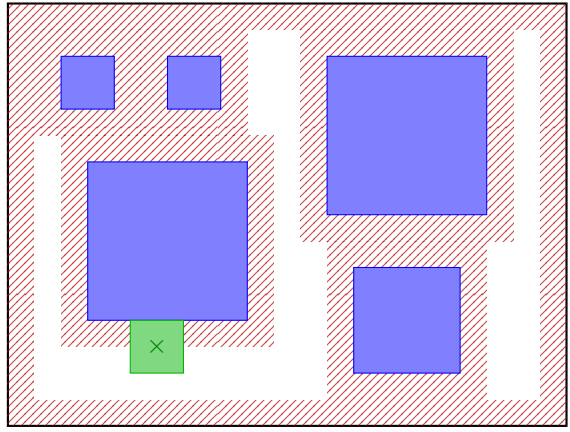
\begin{figure}[t]
 		\centering
 		\begin{tikzpicture}[scale = 0.7, greenrect/.style = {draw = green!70!black, fill = {rgb:green,7;black,3;white,10}}]
 		\draw[thick, black] (0,-0.5) rectangle (10.5,7.5);
 		\draw[draw = none, pattern = north east lines, pattern color = red!70!black, fill opacity = 0.7] (0,-0.5) rectangle (10.5,7.5);
 		\draw[draw = none, fill = white] (0.5,0) rectangle (10,7);
 		\draw[draw = none, pattern = north east lines, pattern color = red!70!black, fill opacity = 0.7] (1,1) rectangle (5,5);
 		\draw[draw = none, pattern = north east lines, pattern color = red!70!black, fill opacity = 0.7] (0.5,5) rectangle (2.5,7);
 		\draw[draw = none, pattern = north east lines, pattern color = red!70!black, fill opacity = 0.7] (2.5,5) rectangle (4.5,7);
 		\draw[draw = none, pattern = north east lines, pattern color = red!70!black, fill opacity = 0.7] (6,0) rectangle (9,3);
 		\draw[draw = none, pattern = north east lines, pattern color = red!70!black, fill opacity = 0.7] (5.5,3) rectangle (9.5,7);
 		\draw[draw = blue, fill = blue!50!white] (1.5,1.5) rectangle (4.5,4.5);
 		\draw[draw = blue, fill = blue!50!white] (1,5.5) rectangle (2,6.5);
 		\draw[draw = blue, fill = blue!50!white] (3,5.5) rectangle (4,6.5);
 		\draw[draw = blue, fill = blue!50!white] (6.5,0.5) rectangle (8.5,2.5);
 		\draw[draw = blue, fill = blue!50!white] (6,3.5) rectangle (9,6.5);
 		\draw[greenrect] (2.3,0.5) rectangle (3.3,1.5);
 		\node at (2.8,1){\textcolor{green!50!black}{$\times$}};
 		\end{tikzpicture}
 		\caption{The white area that is not shaded remains feasible for midpoints of whitespace squares.}\label{FigFeasibleMidpoints}
 	\end{figure}
 	In particular, as the area of the (original) whitespace is at least $F-1$, the area of the set $M$ of potential midpoints of whitespace squares amounts to \[\textrm{area}(M)\geq F-1-\sum_{i=1}^{n}\textrm{area}(F_i)-\textrm{area}(F_R)-\sum_{i=n+1}^{k-1}\mathrm{area}(S'_i).\] We have
 	\[\textrm{area}(F_i)=4\cdot s_i\cdot\frac{s_k}{2}+4\cdot\left(\frac{s_k}{2}\right)^2 = 2\cdot s_k\cdot s_i +s_k^2\] and \[\textrm{area}(F_R)=2\cdot (H+W)\cdot\frac{s_k}{2}-4\cdot\left(\frac{s_k}{2}\right)^2=(H+W)\cdot s_k-s_k^2.\]  By the inequality between arithmetic and quadratic mean, we know that \[
 	\frac{\sum_{i=1}^{n}s_i}{n}\leq \sqrt{\frac{\sum_{i=1}^{n}s_i^2}{n}}\leq\sqrt{\frac{1}{n}} \qquad\Rightarrow\qquad \sum_{i=1}^{n}s_i \leq \sqrt{n}.\]
 	We further have $H=\frac{F}{W}$ and $W\in[\frac{1}{10},\sqrt{F}]$ since $\frac{1}{10}\leq W\leq H$, so by convexity of the function $[\frac{1}{10},\sqrt{F}]\rightarrow\mathbb{R}^+,W\mapsto W + \frac{F}{W}$, which attains its minimum at $\sqrt{F}$ and is, therefore, monotonically decreasing, it follows that \[W+H\leq \frac{1}{10}+\dfrac{F}{\frac{1}{10}} = 10F+ \frac{1}{10}\leq\sqrt{n},\]
 	where the last inequality is implied by our choice of $n$.
 	Moreover, \[\sum_{i=n+1}^{k-1}\mathrm{area}(S'_i) = \sum_{i=n+1}^{k-1} (s_i+s_k)^2\leq \sum_{i=n+1}^{k-1} (2s_i)^2 = 4\sum_{i=n+1}^{k-1} s_i^2\leq 4(c^2-s_k^2),\] because the sequence $(s_i)_{i\geq n+1}$ is non-increasing.
 	Hence, we obtain \begin{align*}\textrm{area}(M) &\geq F-1-\sum_{i=1}^{n}(2\cdot s_i\cdot s_k + s_k^2) -( (W+H)\cdot s_k -s_k^2 )-4(c^2-s_k^2)\\
 	&\geq F-1 -2\sqrt{n}\cdot s_k-n\cdot s_k^2 -\sqrt{n}\cdot s_k + s_k^2 -4c^2+4s_k^2 \\
 	&> F-1-4c^2 -3\sqrt{n}\cdot s_k-n\cdot s_k^2
 	\end{align*}
 	and we need to see that the latter term is non-negative whenever $s_k>0$. To this end, we first want to show that our choice of $c$ implies \begin{equation}4\cdot c^2 \leq F.\label{Fgeq4csquared}\end{equation}
 	We have
 	\begin{alignat}{6}
 	&4\cdot c^2 &&&\leq&\quad F \\
 	\Leftrightarrow\quad& 4\cdot \Biggl( \sqrt{\left(\frac{3}{10}\right)^2+\frac{F-1}{5}}&-&\frac{3}{10}\Biggr)^2 &\leq&\quad F\notag\\
 	\Leftrightarrow\quad& 4\cdot\left(\left(\frac{3}{10}\right)^2+\frac{F-1}{5}\right)&-&\quad 8\cdot \sqrt{\left(\frac{3}{10}\right)^2+\frac{F-1}{5}}\cdot\frac{3}{10}+4\cdot\left(\frac{3}{10}\right)^2 &\leq&\quad F\notag\\
 	\Leftrightarrow\quad& \frac{9}{25}+\frac{4}{5}\cdot F-\frac{4}{5}+\frac{9}{25} &\leq&\quad F+
 	\frac{12}{5}\cdot \sqrt{\left(\frac{3}{10}\right)^2+\frac{F-1}{5}}&&\notag\\
 	\Leftrightarrow\quad& \frac{18}{25}-\frac{20}{25} &\leq&\quad \frac{F}{5}+\frac{12}{5}\cdot \sqrt{\left(\frac{3}{10}\right)^2+\frac{F-1}{5}}&&\notag\\
 	\Leftrightarrow\quad& -\frac{2}{25} &\leq&\quad \frac{F}{5}+\frac{12}{5}\cdot \sqrt{\left(\frac{3}{10}\right)^2+\frac{F-1}{5}},&&\notag
 	\end{alignat}
 	which is true since the left hand side is negative and the right hand side is positive as $F>1$.\\
 	Now, we show that $0<s_k\leq \frac{c}{\sqrt{n}}$ and our assumptions on $c$ and $n$ imply that \[F-1-4c^2 -3\sqrt{n}\cdot s_k-n\cdot s_k^2\geq 0.\]
 	As $F-1-4c^2-3\sqrt{n}\cdot x-n\cdot x^2 = 0$ if and only if \[x=\frac{1}{\sqrt{n}}\cdot\left(-\frac{3}{2}-\sqrt{\frac{9}{4}+F-1-4c^2}\right)\text{ or }x=\frac{1}{\sqrt{n}}\cdot\left(-\frac{3}{2}+\sqrt{\frac{9}{4}+F-1-4c^2}\right),\] where the discriminant is positive by~\eqref{Fgeq4csquared}, we know that $\mathrm{area}(M)>0$, provided that \linebreak[4] $s_k\leq \frac{1}{\sqrt{n}}\cdot\left(-\frac{3}{2}+\sqrt{\frac{9}{4}+F-1-4c^2}\right)$. To this end, as $s_k\leq\frac{c}{\sqrt{n}}$, it suffices to see that we have $c\leq -\frac{3}{2}+\sqrt{\frac{9}{4}+F-1-4c^2}$, or, equivalently, $c+\frac{3}{2}\leq \sqrt{\frac{9}{4}+F-1-4c^2}$ respectively $(c+\frac{3}{2})^2\leq \frac{9}{4}+F-1-4c^2$ since $c>0$.
 	We get \begin{alignat*}{3}
 	&\qquad\left(c+\frac{3}{2}\right)^2\quad &\leq&\quad \frac{9}{4}+F-1-4c^2\\
 	\Leftrightarrow &\qquad c^2 + 3c + \frac{9}{4} &\leq&\quad  \frac{9}{4}+F-1-4c^2\\
 	\Leftrightarrow &\qquad 5c^2 + 3c -(F-1) \quad &\leq&\qquad 0\\
 	\Leftrightarrow &\qquad c^2+\frac{3}{5}c-\frac{F-1}{5} \quad &\leq&\qquad 0,
 	\end{alignat*}
 	which is equivalent to $-\frac{3}{10}-\sqrt{\left(\frac{3}{10}\right)^2+\frac{F-1}{5}}\leq c\leq -\frac{3}{10}+\sqrt{\left(\frac{3}{10}\right)^2+\frac{F-1}{5}}$ and follows from our choice of $c$ and since the left hand term is negative.\\
 	Hence, whenever $s_k>0$, the set of feasible midpoint locations for the respective square is non-empty, which concludes the proof.
 \end{proof}
\begin{lemma}
 Let $c$ be defined as in Lemma~\ref{PackRemainingSquares} and let \[\delta:=\dfrac{F-1}{\frac{10F}{c^2}+\frac{1}{10}}.\] Then any set of squares with total area $V$ satisfying  $c^2\leq V\leq 1$ and maximum edge length at most $\delta$ can be packed into any rectangle of area $F\cdot V$ with greater edge length at most $10F$.
 	\label{delta}
 \end{lemma}
 \begin{proof}
 Let $0<a_1\leq a_2\leq 10F$ be the edge lengths of a rectangle of area $F\cdot V=a_1\cdot a_2$. Then we have $\sqrt{FV}\leq a_2\leq 10F$.
 As the function $(0,\infty)\rightarrow \mathbb{R}^+,$ $x\mapsto x +\frac{FV}{x}$ is convex and attains its minimum at $\sqrt{FV}$, it is monotonically increasing on $[\sqrt{FV},\infty)$. Consequently, we obtain \[a_1+a_2=a_2+\frac{FV}{a_2}\leq 10F+\frac{FV}{10F}=10F+\frac{V}{10}.\] By Corollary~\ref{circumference}, we know that any set of squares with total area $V$ and maximum edge length at most \[\dfrac{(F-1)V}{10F+\frac{V}{10}}=\dfrac{F-1}{\frac{10F}{V}+\frac{1}{10}}\] can be packed into the given rectangle. Since $V\geq c^2$, we have \[\delta=\dfrac{F-1}{\frac{10F}{c^2}+\frac{1}{10}}\leq \dfrac{F-1}{\frac{10F}{V}+\frac{1}{10}},\] so the latter holds in particular for any set of squares of total area $V$ and maximum edge length at most $\delta$. 
 \end{proof}

 \begin{remark}
 A slightly better value of $\delta$ can be achieved in the following way:
 	Let \[K:=\left\{(V,H): c^2\leq V\leq 1, \sqrt{FV}\leq H\leq 10F, \text{\scriptsize$\left(\dfrac{H+\frac{FV}{H}}{4}\right)^2\geq \frac{(F-1)V}{2}$}\right\}.\] Then $K$ is compact.
 	Consider \begin{align*}f:=K&\rightarrow\mathbb{R}^+,\\ (V,H)&\mapsto \dfrac{H+\frac{FV}{H}}{4} - \sqrt{\left(\dfrac{H+\frac{FV}{H}}{4}\right)^2-\frac{(F-1)V}{2}}.\end{align*} This is a continuous function mapping $(V,H)$ to the smallest solution $x$ of \[x^2+\Bigl(H-x\Bigr)\cdot\left(\frac{FV}{H}-x\right)=V,\] if exists. Note that $K$ is just the set of all possible areas $V$ and greater edge lengths $H$ for which the discriminant is non-negative and such a solution does exist. Moreover, observe that we have $x^2+(H-x)\cdot(\frac{FV}{H}-x)\geq V$ for both $(V,H)\not\in K$ or $(V,H)\in K$ and $0\leq x\leq f(V,H)$ since $0^2+(H-0)\cdot(\frac{FV}{H}-0)=FV > V$ and by the intermediate value theorem. If $K$ is non-empty, then by compactness of $K$ and continuity of $f$, let $\delta_1$ be the minimum of $f$, otherwise let $\delta_1=1$. In either case, $\delta_1>0$ since all function values of $f$ are positive because \[\frac{(F-1)V}{2}>0 \quad\text{and}\quad \dfrac{H+\frac{FV}{H}}{4}>0\quad\text{for all positive values of $H$.}\] Let $\delta:=\min\{\delta_1,\frac{c^2}{10}\}$. Then for any $c^2\leq V\leq 1$ and any rectangle $R$ of area $FV$ and with greater edge length $H\leq 10F$, the smaller edge length $W=\frac{FV}{H}$ measures at least $\frac{c^2}{10}\geq \delta$. Besides, by choice of $\delta_1$ and as $\delta\leq \delta_1$, we have $x^2+(H-x)\cdot(W-x)\geq V$ for all $0\leq x\leq \delta$. By Theorem~\ref{MeirMoser}, any set of squares with total area $V$ and maximum edge length at most $\delta$ can be packed into $R$.\hfill $\qedsymbol$ \end{remark}
 \begin{lemma}
 Let $c$ be as in Lemma~\ref{PackRemainingSquares}, $\delta$ as in Lemma~\ref{delta} and define $N_0:=\max\{1,\lfloor\frac{1}{\delta^2}\rfloor\}$. Assume that for any set $S$ of at most $N_0$ squares of total area $1$, there exists a rectangle $R$ of area $F$ into which the squares in $S$ can be packed axis-parallel and internally disjoint.\\
 Let $(s_i)_{i\in\mathbb{N}^+}$ be a non-increasing sequence of non-negative numbers such that\begin{itemize} \item $s_1\geq\frac{1}{10}$, \item $\sum_{i=1}^\infty s_i^2=1$ and \item $\sum_{i=N_0+1}^{\infty}s_i^2\geq c^2$.\end{itemize} Then, there exists a rectangle $R$ of area $F$ into which the squares with edge lengths given by $(s_i)_{i\in\mathbb{N}^+}$ permit an axis-parallel, internally disjoint packing.
 	\label{N0}
 \end{lemma}
 \begin{proof}
	Let $(s_i)_{i\in\mathbb{N}^+}$ be as in the lemma. Note that since $s_1\geq \frac{1}{10}$, the total area of the first $N_0$ squares is strictly positive. Apply the assumption stated for sets of at most $N_0$ squares to the set of squares the edge lengths of which arise from the sequence $(s_i)_{i=1}^{N_0}$ by normalizing the area $\sum_{i=1}^{N_0} s_i^2$ to $1$, and then scale back. We obtain a rectangle $R'$ of area $F\cdot \sum_{i=1}^{N_0}s_i^2 \leq F$ into which we can pack the squares with edge lengths $s_1,\dots, s_{N_0}$. Let $W'$ be the smaller edge length of $R'$. Then \[\frac{1}{10}\leq s_1\leq W'\leq \sqrt{F}\cdot\sqrt{\sum_{i=1}^{N_0}s_i^2}\leq \sqrt{F}.\] Let $V:= \sum_{i=N_0+1}^{\mathbb{\infty}}s_i^2$ and let $R''$ be the rectangle with edge lengths $W'$ and $\frac{FV}{W'}$. Then \mbox{$c^2 \leq V \leq 1$} and for the greater edge length $H''$ of $R''$, we have \[H''=\max\left\{W',\frac{FV}{W'}\right\}\leq\max\left\{\sqrt{F},\dfrac{F}{\frac{1}{10}}\right\}=10F.\]
	Additionally, $s_1\geq\dots\geq s_{N_0}\geq s_{N_0+1}$ and $\sum_{i=1}^{\infty}s_i^2=1$ yield $s_{N_0+1}\leq\frac{1}{\sqrt{N_0+1}}\leq \delta$, so by Lemma~\ref{delta}, we can pack the squares with edge lengths $s_{N_0+1},s_{N_0+2},\dots$ into the rectangle $R''$. Gluing $R'$ and $R''$ together at the edge of length $W'$ yields a rectangle $R$ of area \[F\cdot \left(\sum_{i=1}^{N_0}s_i^2 +\sum_{i=N_0+1}^{\mathbb{\infty}}s_i^2\right)=F\] containing all squares (see Figure~\ref{FigGlueingRectangles}).
\end{proof}
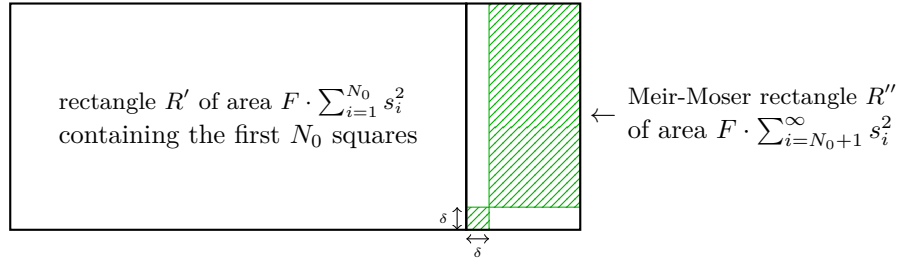
\begin{figure}[t]
\centering
	\begin{tikzpicture}[scale = 0.3]
	\draw[thick] (0,0) rectangle (20,10);
	\node[align = left] at (10,5){\small rectangle $R'$ of area $F\cdot\sum_{i=1}^{N_0}s_i^2$\\ containing the first $N_0$ squares};
	\draw[draw =green!70!black, pattern = north east lines, pattern color = green!70!black] (20,0) rectangle (21,1);
	\draw[<->] (21,-0.4)to (20,-0.4);
	\node at (20.5,-1){\tiny $\delta$};
	\draw[<->] (19.6,0) to (19.6,1);
	\node at (19,0.5){\tiny $\delta$};
	\draw[draw =green!70!black, pattern = north east lines, pattern color = green!70!black] (21,1) rectangle (25,10);
	\draw[thick] (20,0) rectangle (25,10);
	\node at (26,5){$\leftarrow$};
	\node[align = left] at (33,5) {\small Meir-Moser rectangle $R''$\\ of area $F\cdot\sum_{i= N_0+1}^\infty s_i^2$};
	\end{tikzpicture}
	\caption{Gluing together $R'$ and $R''$ provides the desired packing.}\label{FigGlueingRectangles}
\end{figure}
 \begin{theorem}
 Let $N_0$ as in Lemma~\ref{N0} and $c$ as in Lemma~\ref{PackRemainingSquares} and define \begin{itemize}\item $N_1:=\left\lfloor\max\left\{N_0, \left(10F+\frac{1}{10}\right)^2, 100c^2\right\}\right\rfloor$ as well as \item $N:=\lfloor\mathrm{e}^2\cdot N_1\rfloor.$\end{itemize}
 	Then any set of squares of total area $1$ can be packed into some rectangle of total area $F$, provided this holds for any set of at most $N$ squares of total area $1$.
 \end{theorem}
 \begin{proof}
 Note that for any natural number $n$, we have 
 \begin{align*}
  \ln(n+1)&=\int_1^{n+1}\frac{1}{x}dx \leq \sum_{i=1}^{n} (i+1-i)\cdot\frac{1}{i}=\sum_{i=1}^n \frac{1}{i} = 1+\sum_{i=2}^n\frac{1}{i}\\
  &=1+\sum_{i=1}^{n-1}(i+1-i)\cdot\frac{1}{i+1}\leq 1+\int_1^n\frac{1}{x}dx = \ln(n)+1\\ \Rightarrow \ln(n+1)&\leq \sum_{i=1}^n \frac{1}{i}\leq \ln(n)+1.\end{align*}
 By our choice of $N$ and since $N_1\geq\lfloor N_0\rfloor=N_0\geq 1$, we have \begin{align*}\sum_{i=N_1+1}^{N}\frac{1}{i} &= \sum_{i=1}^{N}\frac{1}{i}-\sum_{i=1}^{N_1}\frac{1}{i}\geq \ln(N+1)-\ln(N_1)-1\geq \ln(\mathrm{e}^2N_1)-\ln(N_1)-1\\
 &=\ln\left(\frac{\mathrm{e}^2N_1}{N_1}\right)-1=\ln(\mathrm{e}^2)-1=2-1=1.\end{align*}
 In particular, we get \begin{equation}
                 \sum_{i=N_1+1}^{N}\frac{c^2}{i}\geq c^2. \label{Eqatleastc2}
                \end{equation}
Consider any set of squares of total area $1$ and the corresponding non-increasing sequence $(s_i)_{i\in\mathbb{N}^+}$ of edge lengths.
 	If $s_1\leq\frac{1}{10}$, we are done by Proposition~\ref{small_s1}.\\ So assume $s_1> \frac{1}{10}$. If $\sum_{i=N_1+1}^{\infty}s_i^2 \geq c^2$, then we are done by Lemma~\ref{N0} and our assumption since $N_0\leq N_1$ because $N_0$ is integral, and, therefore, also  $\sum_{i=N_0+1}^{\infty}s_i^2 \geq c^2$.
 	Hence, we can further assume that $\sum_{i=N_1+1}^{\infty}s_i^2 < c^2$. As $\sum_{i=N_1+1}^{N}\frac{c^2}{i} \geq c^2$ by \eqref{Eqatleastc2}, we cannot have $s_i\geq \frac{c}{\sqrt{i}}$ for all $N_1+1\leq i\leq N$, so there is $N_1+1\leq n\leq N$ with $s_n< \frac{c}{\sqrt{n}}$. By our packing assumption for sets of at most $N\geq n$ squares, we know that the squares with edge lengths $s_1,\dots,s_{n-1},\sqrt{1-\sum_{i=1}^{n-1}s_i^2}\geq s_n$ can be packed into some rectangle of area $F$, so this in particular also holds for the squares with edge lengths $s_1,\dots,s_n$.
 	But now, as we also have $\sum_{i=n+1}^{\infty}s_i^2 \leq \sum_{i=N_1+1}^{\infty}s_i^2 < c^2$ and $\frac{c}{\sqrt{n}}\geq s_n\geq s_{n+1}\geq s_{n+2}\geq\dots$, we can apply Lemma~\ref{PackRemainingSquares}, knowing that $n\geq N_1+1\geq\max\{(10F+\frac{1}{10})^2, 100c^2\}$, to conclude the statement of the theorem.
 \end{proof}
 \section{Computing the Value of \texorpdfstring{$N$}{N}}
 In this section, we first compute the value of $N$ we achieve for $F=\frac{2+\sqrt{3}}{3}$ and then comment on some strategies to improve on the obtained value for any $F\geq \frac{2+\sqrt{3}}{3}$. Anyhow, as none of these seems to result in a value of $N$ providing hope to tackle the proof for at most $N$ squares by a computer-aided case distinction, we do not go into full details.\\
 For $F=\frac{2+\sqrt{3}}{3}$, we first get a value of \[c=\sqrt{\left(\frac{3}{10}\right)^2+\frac{F-1}{5}}-\frac{3}{10}\approx 0.07256.\]
 From this, we get \[\delta=\dfrac{F-1}{\frac{10F}{c^2}+\frac{1}{10}}\approx 1.03278\cdot 10^{-4},\] which in turn provides a value of \[N_0=\left\lfloor\frac{1}{\delta^2}\right\rfloor = 93,752,341.\] From this, we obtain \[N_1=\left\lfloor\max\left\{N_0,\left(10F+\frac{1}{10}\right)^2,100c^2\right\}\right\rfloor=N_0=93,752,341\] and, finally, \[N=\left\lfloor \mathrm{e}^2 N_1\right\rfloor=692,741,307.\]
 Observe that this value of $N$ indeed provides an upper bound on the values $N(F)$ for $F\in[\frac{2+\sqrt{3}}{3},1.37]$: First, $c$ is monotonically increasing as a function of $F$, and \[\delta=\dfrac{F-1}{\frac{10F}{c^2}+\frac{1}{10}}=\dfrac{1-\frac{1}{F}}{\frac{10}{c^2}+\frac{1}{10F}}\] is monotonically increasing in both $c$ and $F$, when regarding them as independent parameters. This implies that larger values of $F$ result in larger values of $c$ and $\delta$, and hence smaller values of $N_0$. Given that for values of $F\in[\frac{2+\sqrt{3}}{3},1.37]$, the values $\max\left\{\left(10F+\frac{1}{10}\right)^2,100c^2\right\}$ may attain are several orders of magnitude smaller than the value of $N_0$ we receive for $F=\frac{2+\sqrt{3}}{3}$, it follows that $F=\frac{2+\sqrt{3}}{3}$ yields the maximum value of $N_1$ and, hence, $N$.\\
 In order to improve on this value, there are several possible starting points. First, note that instead of choosing $\frac{1}{10}$ as a threshold for the value of $s_1$, we could take the smaller solution of $x^2+(\sqrt{F}-x)^2=1$, which is greater than $\sqrt{F}-1>\frac{1}{10}$. This slightly improves our bounds on the perimeter of the rectangles under consideration and, therefore, results in larger values of $c$ and $\delta$.\\
 Next, note that in the proof of Lemma~\ref{PackRemainingSquares}, the considered packing has to be chosen particularly unlucky for none of the frames around the squares to intersect with another frame (around a square or inside of the enclosing rectangle), another square or the outside. By considering a bottom-left minimal packing (i.e.\ a packing where no square can be shifted to the left or to the bottom while keeping the other squares fixed), for example, a better bound on the area we lose due to the frames and hence a larger value of $c$ can be established. Moreover, we have already discussed how to (slightly) improve the value of $\delta$ obtained in Lemma~\ref{delta}.\\
 However, a step that seems to be particularly wasteful and that we, therefore, consider worth being discussed in a bit more detail is our choice of $N_0$ in Lemma~\ref{N0}:\\ There, we pick $N_0:=\max\{1,\lfloor \frac{1}{\delta^2}\rfloor\}$ in order to ensure that the $N_0+1^\text{st}$ square has an edge length of no more than $\delta$. Although it is true that this is (almost) the minimum choice of $N_0$ that can guarantee this property, it is not hard to see that a much weaker condition would actually be sufficient for our purposes. To this end, observe that in the case where we really have $s_{N_0+1}=\frac{1}{\sqrt{N_0+1}}$, it immediately follows from the way the edge lengths are ordered and the fact that the total area is $1$ that also $s_1=s_2=\dots=s_{N_0}=\frac{1}{\sqrt{N_0+1}}$. In particular, the edge lengths of the first few squares are extremely small compared to the total area of the remaining ones, which is still close to $1$, so by the same \grqq gluing\grqq\space argument as applied in the proof of Lemma~\ref{N0}, we know that we are actually done at a much earlier stage.\\ In this spirit, the goal of the following considerations is to \grqq speed up\grqq\space the descent down to a total area of at most $c^2$ by considering lower bounds on the edge length $s_n$ of the $n^\text{th}$ square depending on the remaining area $\sum_{i=n+1}^\infty s_i^2$ and showing that we are already done after considering the first $n$ squares if these are not met.\\ More precisely, fix some $F\geq\frac{2+\sqrt{3}}{3}$ and consider a continuous, monotonically increasing function $\delta:[c^2,1]\rightarrow(0,1]$ with the following properties: Given an area $V\in[c^2,1]$ and a rectangle $R$ of area $FV$ such that the greater edge length of $R$ measures at most $10F$ (or $\frac{F}{\sqrt{F}-1}$, applying a better threshold for the minimum size of $s_1$, which still satisfies $\max\left\{\sqrt{F},\frac{F}{\sqrt{F}-1}\right\} =\frac{F}{\sqrt{F}-1}$ as required in the proof of Lemma~\ref{N0}), any set of squares of total area $V$ and maximum occurring edge length $\delta(V)$ can be packed into $R$. Analogously to the proof of Lemma~\ref{delta}, one can see that \[[c^2,1]\rightarrow(0,1],V\mapsto \dfrac{F-1}{\frac{10F}{V}+\frac{1}{10}}\] is a feasible choice for $\delta$. Given such a function $\delta$, we show how to obtain an improved value of $N_0$:
 \begin{lemma}
  Let $N_0:=1+\lfloor\int_{c^2}^1\frac{1}{\delta(V)^2}dV\rfloor$. Note that the considered integral exists since $\delta$ is monotonically increasing with minimum value $\delta(c^2)>0$. Assume that for any set $S$ of at most $N_0$ squares of total area $1$, there exists a rectangle $R$ of area $F$ into which the squares in $S$ can be packed axis-parallel and internally disjoint.\\
  Let $(s_i)_{i\in\mathbb{N}^+}$ be a non-increasing sequence of non-negative numbers such that\begin{itemize} \item $s_1\geq\frac{1}{10}$, \item $\sum_{i=1}^\infty s_i^2=1$ and \item $\sum_{i=N_0+1}^{\infty}s_i^2\geq c^2$.\end{itemize} Then, there exists a rectangle $R$ of area $F$ into which the squares with edge lengths given by $(s_i)_{i\in\mathbb{N}^+}$ permit an axis-parallel, internally disjoint packing.
 \end{lemma}
\begin{proof}
 Consider a non-increasing sequence of non-negative numbers $(s_i)_{i\in\mathbb{N}^+}$ satisfying $s_1\geq\frac{1}{10}$, $\sum_{i=1}^\infty s_i^2=1$ and $\sum_{i=N_0+1}^{\infty}s_i^2\geq c^2$. We show that there exists some $1\leq n\leq N_0$ such that $s_n<\delta(\sum_{i=n}^\infty s_i^2)$. Assume that this is not the case. Then in particular $s_n>0$ for all $1\leq n\leq N_0$ since all function values of $\delta$ are positive, and, moreover, $s_n^2\geq \delta(\sum_{i=n}^\infty s_i^2)^2$ and $\frac{1}{s_n^2}\leq\frac{1}{\delta(\sum_{i=n}^\infty s_i^2)^2}$.\\ We have 
 \begin{align*} N_0 &= \sum_{n=1}^{N_0} \frac{s_n^2}{s_n^2}\leq\sum_{n=1}^{N_0}\frac{s_n^2}{\delta(\sum_{i=n}^\infty s_i^2)^2}=\sum_{n=1}^{N_0}\left(\sum_{i=n}^{\infty}s_i^2-\sum_{i=n+1}^\infty s_i^2\right)\cdot\frac{1}{\delta(\sum_{i=n}^\infty s_i^2)^2}\\
 &\leq \sum_{n=1}^{N_0}\int_{\sum_{i=n+1}^\infty s_i^2}^{\sum_{i=n}^\infty s_i^2}\frac{1}{\delta(V)^2}dV = \int_{\sum_{i=N_0+1}^\infty s_i^2}^{\sum_{i=1}^\infty s_i^2} \frac{1}{\delta(V)^2}dV\leq \int_{c^2}^1 \frac{1}{\delta(V)^2}dV<N_0,
 \end{align*}
  a contradiction. Therefore, there exists some $1\leq n\leq N_0$ such that $s_n<\delta(\sum_{i=n}^\infty s_i^2)$.
 If $n=1$, we are done by definition of $\delta$. Otherwise, our requirements on the function $\delta$ allow us to conduct the same argument as in the proof of Lemma~\ref{N0} to obtain the desired packing.
\end{proof}
Using the function $\delta:[c^2,1]\rightarrow(0,1], V\mapsto \dfrac{F-1}{\frac{10F}{V}+\frac{1}{10}}$ for $F=\frac{2+\sqrt{3}}{3}$, we obtain an improved value of \[N_0=1+\left\lfloor\int_{c^2}^1\frac{1}{\delta(V)^2}dV\right\rfloor = 491,225.\] As a consequence, we get \[N_1=\left\lfloor\max\left\{N_0,\left(10F+\frac{1}{10}\right)^2,100c^2\right\}\right\rfloor=N_0=491,225\] and finally \[N=\lfloor \mathrm{e}^2 N_1\rfloor=3,629,689,\] which is already by a factor of more than $190$ smaller than the previous result. As before, it is not hard to see that the value of $N$ we obtain by applying the previous calculations indeed attains its maximum among all $F\in[\frac{2+\sqrt{3}}{3},1.37]$ for $F=\frac{2+\sqrt{3}}{3}$.\\ We remark that by making use of a slightly more sophisticated version of the $\delta$-function, that arises by attaching the rectangle $R''$ from the proof of Lemma~\ref{N0} at the smaller edge length of $R'$, if $V\leq \frac{1}{2}$, and at the greater edge length, if $V\geq \frac{1}{2}$, it is possible to get down to a value of $N\leq 26,000$. However, as we do not think that a reduction of the provided constant in this order of magnitude changes the qualitative statement of this paper, we spare the details of the computation.

\newpage
\bibliography{square_packing_arxiv.bib}

\begin{thebibliography}{10}

\bibitem{Hougardy11}
Stefan Hougardy.
\newblock On packing squares into a rectangle.
\newblock {\em Computational Geometry}, 44:456--463, 2011.
\newblock \href {https://doi.org/10.1016/j.comgeo.2011.05.001}
  {\path{doi:10.1016/j.comgeo.2011.05.001}}.

\bibitem{Ilhan}
Aylin Ilhan.
\newblock Das {Packen von Quadraten in ein Rechteck}.
\newblock Diploma thesis, Universit\"at Bonn, Forschungsinstitut f\"ur Diskrete
  Mathematik, March 2014.

\bibitem{KleitmanKrieger70}
Daniel Kleitman and Michael Krieger.
\newblock Packing squares in rectangles {I}.
\newblock {\em Annals of the New York Academy of Sciences}, 175:253--262, 1970.

\bibitem{KleitmanKrieger75}
Daniel Kleitman and Michael Krieger.
\newblock An optimal bound for two dimensional bin packing.
\newblock {\em 16th Annual Symposium on Foundations of Computer Science}, pages
  163--168, 1975.
\newblock \href {https://doi.org/10.1109/SFCS.1975.6}
  {\path{doi:10.1109/SFCS.1975.6}}.

\bibitem{Liu17}
Zhiheng Liu.
\newblock On continuity properties for infinite rectangle packing.
\newblock {\em arXiv preprint arXiv:1705.02443}, 2017.

\bibitem{martin2002compactness}
Greg Martin.
\newblock Compactness theorems for geometric packings.
\newblock {\em Journal of Combinatorial Theory, Series A}, 97(2):225--238,
  2002.
\newblock \href {https://doi.org/10.1006/jcta.2001.3209}
  {\path{doi:10.1006/jcta.2001.3209}}.

\bibitem{MeirMoser}
Aram Meir and Leo Moser.
\newblock On packing of squares and cubes.
\newblock {\em Journal of Combinatorial Theory}, 5(2):126--134, 1968.
\newblock \href {https://doi.org/10.1016/S0021-9800(68)80047-X}
  {\path{doi:10.1016/S0021-9800(68)80047-X}}.

\bibitem{MoonMoser}
John Moon and Leo Moser.
\newblock Some packing and covering theorems.
\newblock In {\em Colloquium mathematicum}, volume~17, pages 103--110.
  Institute of Mathematics Polish Academy of Sciences, 1967.

\bibitem{MoserProblemFormulation}
Leo Moser.
\newblock Poorly formulated unsolved problems of combinatorial geometry.
\newblock {\em Mimeographed list}, 1966.

\bibitem{Novotny95}
Pavel Novotný.
\newblock A note on a packing of squares.
\newblock {\em Stud. Univ. Transp. Commun. Zilina Math.-Phys. Ser.}, 10:35--39,
  1995.

\bibitem{Novotny96}
Pavel Novotný.
\newblock On packing of squares into a rectangle.
\newblock {\em Archivum Mathematicum (BRNO)}, 32:75--83, 1996.

\bibitem{Novotny99}
Pavel Novotný.
\newblock On packing of four and five squares into a rectangle.
\newblock {\em Note di matematica}, 19/2:199--206, 1999.
\newblock \href {https://doi.org/10.1285/i15900932v19n2p199}
  {\path{doi:10.1285/i15900932v19n2p199}}.

\bibitem{Platz}
Alexander Platz.
\newblock A proof of {Moser's} square packing problem for small instances.
\newblock Master's thesis, Universit\"at Bonn, Forschungsinstitut f\"ur
  Diskrete Mathematik, August 2016.

\bibitem{Zernisch12}
Jan Zernisch.
\newblock A generalization of a theorem of {K}leitman and {K}rieger.
\newblock {\em International Journal of Computational Geometry \&
  Applications}, 22(02):167--185, 2012.
\newblock \href {https://doi.org/10.1142/S0218195912500033}
  {\path{doi:10.1142/S0218195912500033}}.

\end{thebibliography}

\appendix

\end{document}